\documentclass{article}
\usepackage{spconfa4,amsmath,graphicx}
\usepackage{amssymb}
\usepackage{multirow}
\usepackage{enumitem}

\newcommand{\argmax}{\operatornamewithlimits{argmax}}

\title{Multi-label Zero-Shot Audio Classification with Temporal Attention}
\name{Duygu Dogan, Huang Xie, Toni Heittola, Tuomas Virtanen} 
\address{Unit of Computing Sciences, Tampere University, Finland}
\begin{document}
\ninept
\maketitle
\begin{abstract}
Zero-shot learning models are capable of classifying new classes by transferring knowledge from the seen classes using auxiliary information. While most of the existing zero-shot learning methods focused on single-label classification tasks, the present study introduces a method to perform multi-label zero-shot audio classification. To address the challenge of classifying multi-label sounds while generalizing to unseen classes, we adapt temporal attention.~The temporal attention mechanism assigns importance weights to different audio segments based on their acoustic and~semantic~compatibility, thus enabling the model to capture the varying dominance of different sound classes within an audio sample by focusing on the segments most relevant for each class. This leads to more accurate multi-label zero-shot classification than methods employing temporally aggregated acoustic features without weighting, which treat all audio segments equally. We evaluate our approach on a subset of AudioSet against a zero-shot model using uniformly aggregated acoustic features, a zero-rule baseline, and the proposed method in the supervised scenario. Our results show that temporal attention enhances the zero-shot audio classification performance in multi-label scenario.

\end{abstract}
\section{Introduction}
\label{sec:intro}
\vspace*{-0.2\baselineskip}

Audio classification is one of the fundamental tasks in the field of signal processing and machine learning, aiming to categorize audio data into predefined classes. The task has been applied in various domains and is traditionally approached by supervised learning methods. However, despite their efficacy, supervised learning models suffer from certain limitations. One significant limitation is the inability to generalize the model to the classes not introduced during the development stage, inferring that continuous model and dataset maintenance is needed as new classes emerge over time. Furthermore, for these methods to perform robustly, the presence of a large set of annotated data is required and the acquisition of such data is generally difficult and expensive. 

Zero-shot learning (ZSL), proposed in \cite{zsl_org}, is a paradigm that aims to transfer knowledge from seen to unseen classes by leveraging their auxiliary information without explicit training data, thus addressing the limitations of traditional supervised learning. In recent years, zero-shot classification has drawn much attention, particularly within the vision community \cite{zsl_img1, zsl_img2, zsl_img3, zsl_img4}. While ZSL has found applications also in audio-visual learning \cite{audiovisual-9880403, audiovisual-Mercea2022}, its adoption in the general audio domain has been more scattered. ZSL has been studied in specific contexts such as in music genre classification \cite{music}, in speech processing tasks \cite{speech}, and in bioacoustics \cite{10445807}.  Previous research in \cite{islam_zsl_audio, xie_zsl_audio1, xie_zsl_audio2} studied general-purpose zero-shot audio classification by leveraging semantic information of the sound classes using the class labels. The follow-up \cite{xie_zsl_audio3} work also experimented with semantic sentence-level embeddings. Our prior work~\cite{dogan_zsl_audio} explored the usage of image semantic embeddings for zero-shot audio classification. However, existing methods proposed for zero-shot audio analysis tasks have been predominantly dedicated to performing single-label audio classification tasks, where each audio instance is assigned to a single category. 

Performing multi-label classification in a zero-shot learning setup is challenging because typical ZSL methods, which calculate similarity between semantic and acoustic embeddings, are not designed to recognize multiple overlapping classes. Hence, when the training data is multilabel, the conventional models cannot be properly trained. Yet, in real-world audio scenes, multiple sounds often occur together within an audio sample, requiring the development of multi-label learning approaches to meet the complexity of the audio signals. Multi-label zero-shot learning (ML-ZSL) in the audio domain has been studied in~\cite{music} for music tagging and in the general purpose audio classification in~\cite{passt}. The authors in \cite{passt} explored the usage of Patchout Spectrogram Transformers as the audio encoder to extract the acoustic embeddings. However, their approach aggregates the extracted acoustic embeddings over time, overlooking the fact that oftentimes sound events are not active throughout the whole audio sample and different sounds mainly dominate different parts of the audio. 

In our work, we study multi-label zero-shot learning in the general-purpose audio classification by utilizing the temporal information of audio data. Rather than aggregating the acoustic features uniformly over time, we aim to develop a model that gives different importance weights to each audio segment of an audio sample. For this purpose, we adapt the temporal attention mechanism. Temporal attention has often been used with recurrent \cite{attn1} and convolutional neural network \cite{attn2} layers to focus on different parts of the input sequence at each step. It has also been explored in audiovisual learning to model the multimodal dependencies \cite{audiovisual-Mercea2022}. In this work, we employ temporal attention on the compatibility between the acoustic and semantic features to model the variability of the temporal activity of sound events in an audio sample. The attention module allows the model to focus on the audio segments most relevant for each class, hence enabling multi-label classification.  We compare our results with i) a baseline model that uses uniformly aggregated features, ii) a zero-rule baseline where the predicted class is always the most frequent one, and iii) the proposed method in the supervised classification setup.

The remainder of this paper is organized as follows: Section~\ref{sec:method} presents our proposed method. Section \ref{sec:experiments} discusses experimental setup including dataset and evaluation metrics. Section \ref{sec:results} reports the experimental results with comparative analysis. Finally, Section \ref{sec:conclusion} concludes the paper with key findings and discussion.

\begin{figure*}[ht]
\centering
\vspace*{-1cm}
    \centering
    \includegraphics[]{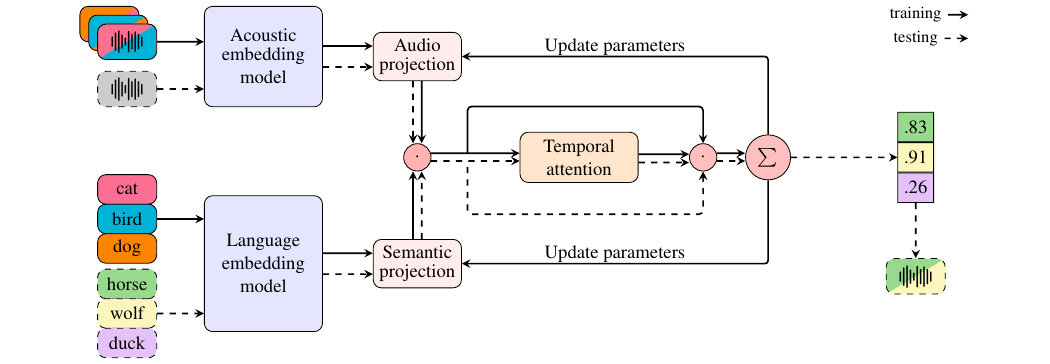}
    \caption{Illustration of the proposed multi-label zero-shot model. Training in solid and testing in dashed line. Acoustic and semantic embeddings are extracted from audio samples and class labels using pre-trained embedding models. The embeddings are projected into a common space, combined using a compatibility measure, and weighted by a temporal attention module. The weighted scores are aggregated to produce final prediction scores. During testing, model processes unknown audio samples and new class labels, predicts the classes with highest scores.}
    \label{fig:pipeline}
\end{figure*}

\section{Proposed Method}
\label{sec:method}
In this section, we formulate the problem setup and discuss our proposed method for multi-label zero-shot audio classification settings. 
\vspace*{-0.42cm}
\subsection{Problem setting}
The pipeline of the proposed model is depicted in Figure~\ref{fig:pipeline}. Acoustic and semantic embeddings are extracted from audio samples and class labels using pre-trained acoustic and language embedding models. These embeddings are projected into a common dimensional space through audio and semantic projection layers. A compatibility measure combines the projected embeddings to compute similarity scores. A temporal attention module weights these scores to determine relevant segments. The weighted scores are aggregated to obtain final prediction scores for each class. During testing, the model processes unknown audio samples and new class labels and outputs predicted labels determined by the highest prediction scores.

We denote the audio space by $X$, and training and zero-shot sound classes by $Y$ and $Z$, respectively, where $Y \cap Z=\emptyset$. We are given a training set $D^{tr} = \{(x_i, S_i) \ | x_i \in X, S_i  \subseteq Y\}$, and $D^{test} = \{(x_i, U_i) \ | x_i \in X, U_i  \subseteq Z\}$, where $x_i$ is an audio sample and $S_i$ and $U_i$ are the set of corresponding clip-level labels belonging to training and zero-shot sound classes, respectively. Our aim is to obtain a classifier that recognize the multiple unseen classes in~$U~\subseteq~Z$ of an audio sample. Using an acoustic embedding model $\theta$ and a semantic embedding model $\phi$, we denote the acoustic embeddings of an audio sample~$x$ by $\theta(x) \in \mathbb{R}^{F_a \times T}$ and the semantic embeddings of a class $y_c$ by $\phi(y_c) \in \mathbb{R}^{F_s}$. Here, an audio sample is the entire audio recording. Each sample is divided into smaller, fixed-length time units and called segments. The acoustic embedding model ~$\theta$ maps these audio samples to acoustic embeddings, where $T$ is the number of segments and $F_a$ is the dimensionality of the acoustic embeddings extracted independently in each segment. The semantic model $\phi$ is a language embedding model that maps class labels to their corresponding semantic embeddings, with $F_s$ being the number of semantic features. 

\subsection{Model architecture}
In our proposed model, we project semantic and acoustic embeddings into the same shared space where the compatibility between them is calculated. We define the audio projection as $f_a: \mathbb{R}^{F_a \times T} \xrightarrow{} \mathbb{R}^{D_\text{model} \times T}$ and semantic projection as~$f_s:~\mathbb{R}^{F_s}\xrightarrow{}\mathbb{R}^{D_\text{model}}$:
\begin{gather}
    \mathbf{v_x} = f_{a}(\theta(x)) \label{eq:eq1} \\
    \mathbf{v_c} = f_{s}(\phi(y_c)), \label{eq:eq2} 
\end{gather}
where $\mathbf{v_x}^{D_\text{model} \times T}$ is the projected audio embeddings for audio sample $x$ and $\mathbf{v_c}^{D_\text{model}}$ is the projected semantic embedding for the class label $y_c$. Both $f_a$ and $f_s$ can be implemented as a stack of fully connected layers. Then, we obtain the compatibility between $\mathbf{v_c}$ and $\mathbf{v_x}$  using a compatibility function $g$ as
\begin{equation}
    \mathbf{s_c}^\top = g(\mathbf{v_x},\mathbf{v_c}), \label{eq:eq3}
\end{equation}
where we define $g(a,b) = b^\top a$. The resulting vector $\mathbf{s_c} \in \mathbb{R}^T$ contains the similarity values between the semantic embedding of the given class label $y_c$ and the acoustic representations of each time segment. While the compatibility function is the same as in single-label techniques, modeling multi-label data where different classes dominate different segments requires a more advanced approach. Temporal attention allows learning multiple classes within an audio sample by estimating their temporal attention weights. We normalize $\mathbf{s}$ using softmax over the $T$ segments for each $t \in [1, T]$ as
\begin{equation}
    \alpha_{c,t} = \dfrac{\text{exp}(s_{c,t})}{\sum\nolimits_{t' = 1}^T \text{exp}(s_{c,t'})}. \label{eq:eq4} 
\end{equation}
The attention weights $\alpha_{c,t}$ determine the importance of each segment $t$ for each class. We use $\alpha_{c,t}$ to weigh the $s_{c,t}$ and then aggregate over time to obtain prediction scores of each class: 
\begin{equation}
    \hat{y}_c = \sum\nolimits_{t = 1}^{T}{\alpha}_{c,t} s_{c,t}.\label{eq:eq5}
\end{equation}
At test time, we also use the temporally aggregated acoustic features without weighting to make clip-level prediction $\hat{z}_c$ for each class $c$: 
\begin{equation}
    \hat{z}_c = \sum\nolimits_{t = 1}^T s_{c,t}. \label{eq:eq6}
\end{equation}
To capture both global and local features, we obtain the prediction score $\hat{p}$ of a sound event class by fusing $\hat{y}$ and $\hat{z}$ as
\begin{equation}
    \hat{p}_c = \gamma \hat{y}_c + (1 - \gamma) \hat{z}_c. \label{eq:eq7}
\end{equation}
The predicted classes $\hat{C}$ as the set of classes whose prediction score is greater than threshold $M$. 
\begin{equation}
    \hat{C} = \{c \ | \ \hat{p}_c > M\}, 
\end{equation}
where $\gamma$ and $M$ are hyperparameters chosen empirically during validation and set to 0.5 and 1.51, respectively. 

\subsection{Loss function}
During training, the goal is to optimize the weights of the layers in $f_a$ and $f_s$ in such a manner that given a training sample $(x_n, y_n)$, $n$~=~$\argmax_c{\hat{y}_c}$. For this purpose, we consider hinge loss $l$:
\begin{gather}
    l(x, y_+, y_-) = \text{max}\{0,\Delta + \hat{y}_- - \hat{y}_+\},
\end{gather}
where $y_+$ and $y_-$ are respectively the positive and negative classes for given audio sample $x$ and $\Delta$ is a hyperparameter that is set to 1. We use a ranking penalty to consider the loss values in a weighted manner giving greater importance to classes with higher predicted scores. Given a list $\mathbf{y} = (y_{c_i} \ | \ i = {1, \dots |Y|})$ which values are sorted in descending order, we denote $i_{y_c}$ as the index of $y_c$ in $\mathbf{y}$.
The ranking penalty $\beta(i_{y_c})$ is defined as 
\begin{gather}
    \beta(i_{y_c}) = \sum\nolimits_{j=1}^{i_{y_c}} \dfrac{1}{j}, \quad \text{and } \beta(0) = 0.
\end{gather}
The final objective function is weighted approximate-rank pairwise loss \cite{warp_loss} and defined as
\begin{gather}
    \dfrac{1}{N} \sum\limits_{n=1}^N \dfrac{\beta(i_{\hat{y}_c})}{i_{\hat{y}_c}} \sum\limits_{y \in Y} l(x, y_+, y_-).
\end{gather}

\section{Experiments}
\label{sec:experiments}
In this section, we describe the dataset, data split for multi-label zero-shot learning setup, acoustic and semantic embeddings, used metrics and evaluation setup, and implementation details.

\subsection{Dataset}
\label{ssec:dataset}

We conducted our experiments based on AudioSet~\cite{audioset}. AudioSet is a general-purpose audio dataset that includes 527 sound classes with over two million audio samples in total. The audio samples in the dataset contain weak labels defined in one or several words and longer sentence descriptions for all the available sound classes. 

The AudioSet Ontology organizes classes hierarchically in a tree-like structure. For instance, a sample might be categorized under "Animal", "Domestic animals, pets", and "Dog", all referring to the same sound. The Ontology includes a "Restrictions" field, marking~56 categories due to being obscure, and 22 as "abstract" that are used as intermediate nodes to build the ontology structure \cite{audioset}. In the tree structure, we use the terms parent and child class to define the hierarchical relation of two class labels. For example, starting from the top node, in the branch of "Animal"~-~"Wild animals" - "Bird" - "Gull, seagull" we refer "Bird" as the child of "Wild animals" and the parent of "Gull, seagull". To simplfy evaluation, we included only one class from each branch of the hierarchical tree in our experiments. The criteria for selecting these classes were: 

1. We omitted all the restricted classes. 

2.~We excluded classes that had multiple parents from different branches to ensure more distinct boundaries between classes, such as "Cowbell" which is classified under both "Animal" and "Musical instrument".

3. From the remaining classes, we mostly selected the class at the deepest node in each branch with a couple of exceptions. Firstly, we kept the parent class if it only had one or two child classes to avoid narrow classification for the categories that do not have enough data samples. Secondly, under the Animal categories, we chose specific animal classes, e.g. "Dog", rather than their specific sounds e.g. "Bark". This was because a significant amount of the animal categories do not have more than two child classes to provide enough samples, and we aimed to maintain clear semantic distinctions even when some animal classes had more than two child classes.

\vspace*{-0.6\baselineskip}
\subsection{Data split for ML-ZSL setup}
\label{ssec:data_split}
In zero-shot learning setups, the training, validation, and test splits are assumed to have disjoint classes. However, defining these seen and unseen classes in a multi-label zero-shot learning setup is more complex compared to single-label zero-shot learning since each data sample is associated with multiple labels in ML-ZSL. To have a similar label count, balanced sample size, and disjoint classes in the training, validation, and test sets, we follow a greedy algorithm that splits the dataset into three sets (fold1, fold2, fold3) by iteratively adding classes. Initially, the three least frequent classes are placed in fold1, fold2, and fold3. The algorithm then repeatedly finds and adds the classes that most frequently coexist with the last added class in each set until no new classes can be added. Table~\ref{tab1} shows the number of labels and samples in each fold.

\begin{table}[ht]
\centering
\begin{tabular}{l|l|l}
\textbf{set} & \textbf{number of classes} & \textbf{number of samples} \\ \hline \hline
fold1 & 111 &  7822\\ 
fold2 & 111 &  9397\\ 
fold3 & 112 &  6521\\ \hline
\end{tabular}
\caption{Number of labels and samples in each split}\label{tab1}
\end{table} 
\vspace*{-1.5\baselineskip}
\subsection{Acoustic and semantic embeddings}
\label{ssec:embeddings}
For acoustic feature extraction, we utilized a pre-trained VGGish model \cite{vggish}. The process involved the following steps: (1) splitting each 10-second audio sample into ten 1-second segments; (2) converting these segments into log-mel spectrograms, each with 64 mel-filter banks and a temporal resolution of 96 time-frames; (3) passing these spectrograms through the VGGish model to obtain 128-dimensional audio embeddings.

To encode the semantic information of classes, we employed the BERT model \cite{devlin2018bert}. Each class label was transformed into a 768-dimensional vector using BERT. For classes consisting of multiple words, we first embed each word and then compute their average to produce a single vector representing the entire class. 

\vspace*{-0.6\baselineskip}
\subsection{Metrics and evaluation setup}
\label{ssec:evaluation}
We employ both micro and macro F-scores to evaluate our multi-label zero-shot audio classification framework.~Micro F-score counts the total true positives, false negatives, and false positives globally and aggregates the contributions of all classes and thus reduces the effects arising from label imbalance. Macro F-score computes the F-score independently for each class and then takes the average, hence considering all classes equally independent of their number of samples. Macro F-score is often more useful to reflect the model performance on the less frequent labels.

The evaluation of our proposed method is structured around three primary experimental setups:

1. Zero-Shot Setup: In this setup, the training, validation, and testing classes are completely disjoint. We implement three different settings to reduce the bias. Table~\ref{tab2} shows the training, validation, and test folds of these settings. We report the model performance using the proposed model with temporal attention mechanism and compare it with the model from \cite{dogan_zsl_audio} that uses aggregated acoustic embeddings, where $\gamma = 0$ in (\ref{eq:eq7}). In \cite{dogan_zsl_audio}, the features are temporally aggregated without weighting, considering each audio segment in equal importance and hence will be referred to as uniform aggregation. For a fair comparison, we modified the model by employing the BERT model \cite{devlin2018bert} as the semantic embedding model and implemented $f_a$ and $f_s$ as described in Section~\ref{ssec:implement}. 

It should be noted that the acoustic embedding model should ideally be trained from scratch to reduce bias from the classes used to train the model. However, both the temporal attention and uniform aggregation methods use the same pre-trained model, ensuring that the comparison between them is under the same conditions.

\begin{table}[ht!]
\centering
\begin{tabular}{l|l|l|l}
\textbf{Setup} & \textbf{training} & \textbf{validation} & \textbf{test}  \\ \hline \hline
Setting 1 & fold2 & fold3 & fold1  \\ 
Setting 2 & fold3 & fold1 & fold2  \\ 
Setting 3 & fold1 & fold2 & fold3  \\ \hline
\end{tabular}
\caption{Zero-shot classification settings}\label{tab2}
\end{table} 
\vspace*{-0.6\baselineskip}

2. Zero-Rule Baseline: As a baseline, we employ a zero-rule approach where the predicted class is always the most frequent one in the test set. This method serves as a lower benchmark.

3. Supervised Classification Setup: For a comparative analysis, we also implement a supervised learning scenario without the constraint of disjoint classes. Here, we utilize cross-validation folds where training, validation, and test classes in this scenario correspond to the same test classes and samples used in each zero-shot experiment setting. This setup is used to set the upper bound of the model performance.

Each model configuration is evaluated using the described setups. The results are averaged over three runs to ensure reliability. 

\subsection{Implementation details}
\label{ssec:implement}
We implemented $f_a$ in (\ref{eq:eq1}) with one fully connected layer with 128 units, followed by a tanh activation and $f_s$ in (\ref{eq:eq2}) with two fully connected layers having 512 and 128 units respectively, each followed by tanh activation. All the models are trained using a batch size of 32, and SGD optimizer with a learning rate of 0.01 for 100 epochs.

\section{Results and Analysis}
\label{sec:results}
In this section, we present the experimental results in multi-label zero-shot classification.

Micro and macro F1-scores are reported in Table~\ref{tab3} and Table~\ref{tab4}, respectively. The tables report the F1-score on the corresponding test fold and their average. The results consist of two zero-shot learning setups; the proposed method using temporal attention and uniform aggregation, a zero-rule baseline as the lower benchmark, and the supervised setup as the measurable upper bound. 

\begin{table}[ht!]
\centering
\begin{tabular}{l | l l | l l }
 test fold & \textbf{\shortstack{Temporal \\ Attention}} & \textbf{\shortstack{Uniform \\ Aggregation}} & \textbf{\shortstack{Zero-rule \\ Baseline}} & \textbf{Supervised} \\ \hline \hline
fold1 & 0.14 & 0.11 & 0.06 & 0.42 \\ 
fold2 & \textbf{0.22} & 0.18 & 0.19 &  0.49 \\ 
fold3 & 0.18 & 0.14 & 0.14  &  0.45 \\ \hline \hline
mean & 0.18 & 0.14 & 0.13 & 0.45\\ \hline
\end{tabular}
\caption{Micro F1-scores}\label{tab3}
\end{table} 

\begin{table}[ht!]
\centering
\begin{tabular}{l | l l | l l}
 test fold & \textbf{\shortstack{Temporal \\ Attention}} & \textbf{\shortstack{Uniform \\ Aggregation}} & \textbf{\shortstack{Zero-rule \\ Baseline}} & \textbf{Supervised} \\ \hline \hline
fold1 &  0.02 & 0.009 & 0.002 & 0.30 \\ 
fold2 & \textbf{0.07}  & 0.010 & 0.007 &  0.36 \\ 
fold3 & 0.04  & 0.006 & 0.003  & 0.31 \\ \hline \hline
mean & 0.04 & 0.008  & 0.004 & 0.32 \\ \hline
\end{tabular}
\caption{Macro F1-scores}\label{tab4}
\end{table} 

The model trained in a supervised setup achieves the highest micro and macro F1-scores across all folds. The temporal attention model outperforms both the model with uniform aggregation and zero-rule baseline in all folds, showing that the usage of temporal attention improves the model's ability to classify unseen classes. The significant difference between the macro F1-scores of the temporal attention and the uniform aggregation indicates that temporal attention has a better capability of handling class imbalances.

\begin{figure}[ht]
  \centering
  \includegraphics[width=0.8\columnwidth]{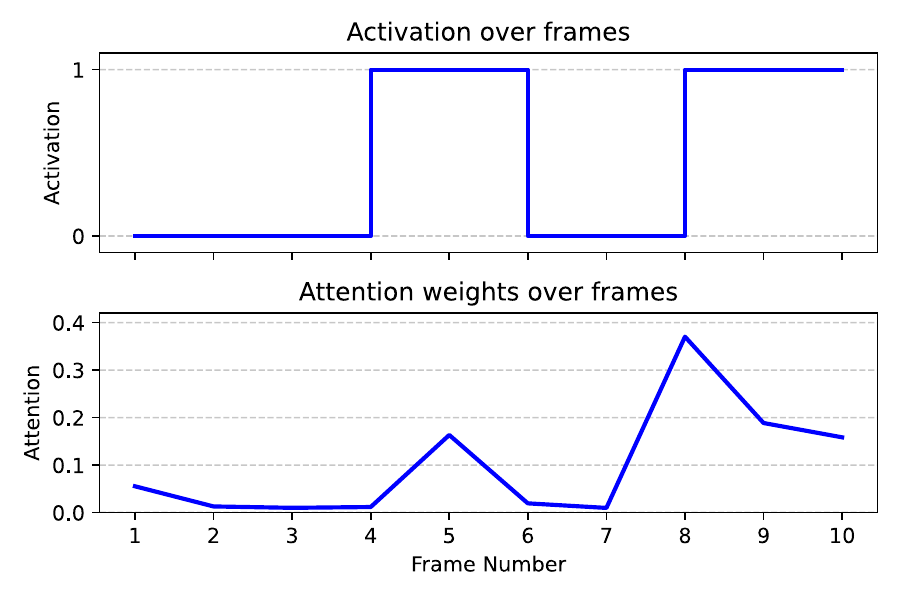}
  \caption{Strong labels (top) and attention weights (bottom) for class label dog of the audio sample "-4gqARaEJE\_0\_10.wav".}
  \label{fig:dog_attn}
\end{figure}
\vspace*{-0.4\baselineskip}
In Figure~\ref{fig:dog_attn}, we demonstrate the strong labels and attention weights over time frames for the class label "dog" of an audio sample. The figure shows that the time frames of the higher attention weights correspond to the frames where the sound event of the "dog" class is active. Even though it is important to note that the figure visualization is only for one sample, it demonstrates the model's potential.

\vspace*{-0.5\baselineskip}
\section{Conclusion}
\label{sec:conclusion}
In this paper, we introduced a temporal attention mechanism for zero-shot audio classification designed to perform multi-label classification by addressing the limitations of clip-level feature aggregation, which overlooks the segment-level information for non-dominant sound events. The semantic information was represented using textual embeddings, and the classification task was conducted on a subset of AudioSet. We evaluated our approach against conventional methods; a baseline model using aggregated acoustic embeddings, a zero-rule baseline, and the supervised classification setup. Our experimental results demonstrated that temporal attention is effective in improving classification performance across all tested classes in multi-label zero-shot scenarios. The results show that temporal attention is a promising approach worth further exploration in zero-shot sound event detection tasks.

\bibliographystyle{IEEEbib}

\end{document}